\documentstyle[11pt,newpasp,twoside,epsf]{article}
\def\edcomment#1{\iffalse\marginpar{\raggedright\sl#1\/}\else\relax\fi}
\reversemarginpar

\begin{document}
\title{Multi-Frequency VSOP and VLBA Polarization Observations of 
3C~279 and 3C~345}
\author{C.~C. Cheung, D.~H. Roberts, J.~F.~C.~Wardle}
\affil{Department of Physics, Brandeis University, Waltham, MA 02454}
%ccheung@brandeis.edu, MS~057,

\author{G.~A. Moellenbrock}
\affil{National Radio Astronomy Observatory, Socorro, NM 87801}
%P.O. Box 0,

\author{D.~C. Homan}
\affil{National Radio Astronomy Observatory, Charlottesville, VA 22903}

\begin{abstract}
This contribution presents preliminary results from coordinated
polarization sensitive VSOP and VLBA imaging of the blazars 3C~279 and
3C~345 at multiple frequencies.
\end{abstract}

\section{Imaging Relativistic Jets with Space-VLBI}

One of the primary goals of orbiting-VLBI, and the VSOP mission in
particular, is to image compact radio sources at low frequencies with
resolution comparable to that of higher frequency ground-based only
observations.  This facilitates resolution independent spectral and
polarization (rotation measure and field orientation) mapping of
relativistic jets in AGN, a capability that is compromised when only
ground-based interferometers are used.

To this end, we have obtained polarization-sensitive observations of four
bright 3C quasars using the VLBA at 8.4, 15, 22, and 43 GHz, plus
coordinated VSOP plus ground station observations at 1.6 and 5 GHz. Here,
we present some preliminary results from our first-epoch mapping of 3C~279
(in April 1999) and 3C~345 (from July-Sept. 1998).  Chen et al. (2003,
this volume) present results on 3C~454.3; work on the fourth source,
3C~273, is in progress.

%and ability that is somewhat 
%when observing with only ground-based interferometers.

\section{The Parsec-scale Jets in 3C~279 and 3C~345}

The quasars 3C~279 (z=0.536) and 3C~345 (z=0.593) are well-studied VLBI
sources (e.g. Piner et al. 2003, Klare et al. 2003)  and are ideal targets
for this study because of their bright pc-scale jets.  Figures~1 \&~2 show
selected images from our first-epoch observations. The 5 GHz VSOP images
of 3C~345 previously appeared in Moellenbrock, Roberts, \& Wardle (2000).

%and shared by Klare et al.
%and refs therein) 

Although the addition of the spacecraft mitigates the resolution issue in
our study, the non-simultaneity (5 days for 3C~279 and $\sim$2 months for
3C~345) of the VLBA only and VSOP observations hampers a direct comparison
of the images because of flux variability; we plan to resolve this by
extrapolating from other contemporaneous data. We aligned the images using
well-defined optically thin features in the jets as labeled in the
figures. The proper motion of C4 in 3C~279 (0.4 mas/yr; Homan et al. 2003)
does not compromise the image registration over the 5 day difference
between observing epochs, and assuming a typical motion of $\sim$0.3
mas/yr (Ros, Zensus, \& Lobanov 2000), the shift in 3C~345 is only
$\sim$1/8$^{\rm th}$ the beamwidth in the jet direction at 5 GHz (Fig.~2).

\begin{figure}
\vspace{-0.59in}
%\plotfiddle{file}{vsize}{rot}{hsf}{usf}{htrans}{utrans}
\plotfiddle{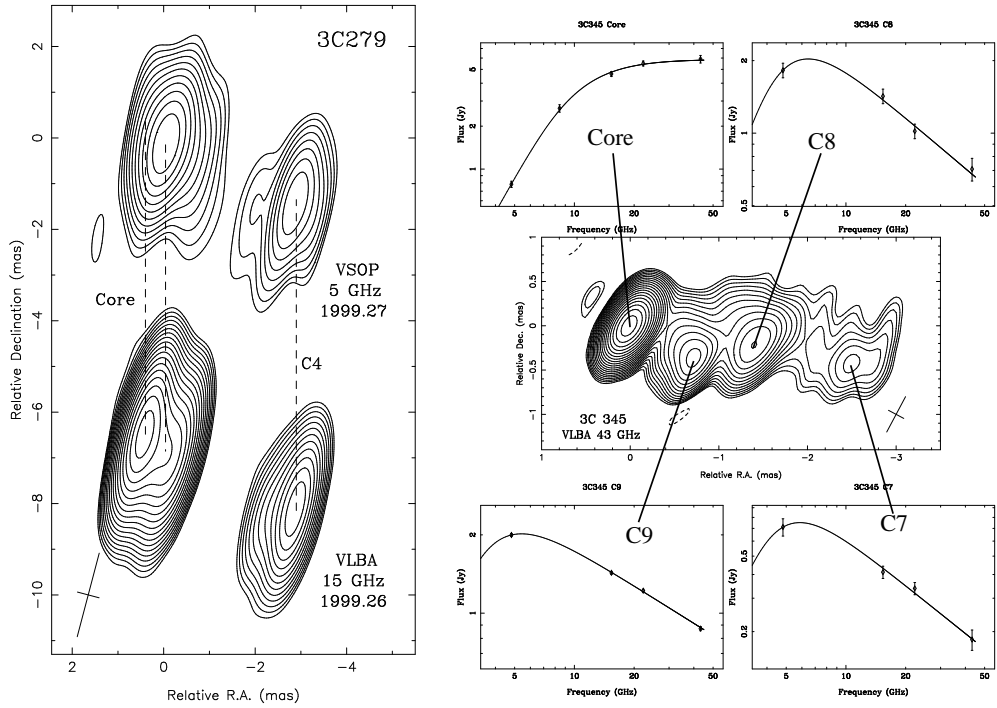}{4.5cm}{0}{59}{59}{-188}{-200}
\vspace{2.65in}
\caption{[left] Matched resolution total intensity images of 3C~279   
(using restoring beam from the VSOP observation).
[right] Spectra of the 3C~345 core and jet
features indicated on a VLBA 43 GHz image. The self-absorbed
spectra fit to the data are shown.
\label{fig1}}
\end{figure}

Once registered, we see in both objects, that the brightest feature at 5
GHz is further downstream from the core than at the higher frequencies,
and is not the core itself (seen in the higher frequency images); thus
opacity effects are important within our observing bands.  In 3C~345, the
modelfits (Fig.~1)  and polarization maps (Fig.~2) show that the core is
almost completely self-absorbed and depolarized at 5 GHz. This may explain
the lack of circular polarization in the core at 8.4 GHz, but its
detection at 15 GHz (Homan \& Wardle 2003).  We find also a low rotation
measure in the core and jet, as measured previously by Taylor (1998).  
The 3C~345 spectra also show hints that the jet is beginning to become
opaque near 5 GHz -- this should become more apparent in the 1.6 GHz VSOP
data. Lastly, there is a noticeable shift in the position of its core
measured at 5 GHz, with respect to the higher frequency 15 to 43 GHz
observations, which can not be reconciled by component motions over the
$\sim$2 month elapse between the observations -- it would require that all
three jet components move more than 3 times faster than typically observed
(Ros et al. 2000).  Analysis of the L-band VSOP data for both objects are
in progress, and observations at other epochs will track variability.

%(the polarized bit may be part of the emerging component C10 seen at
%higher resolution (Klare et al.)  

\begin{figure}
\vspace{-2.2in}
%\plotfiddle{file}{vsize}{rot}{hsf}{usf}{htrans}{utrans}
\plotfiddle{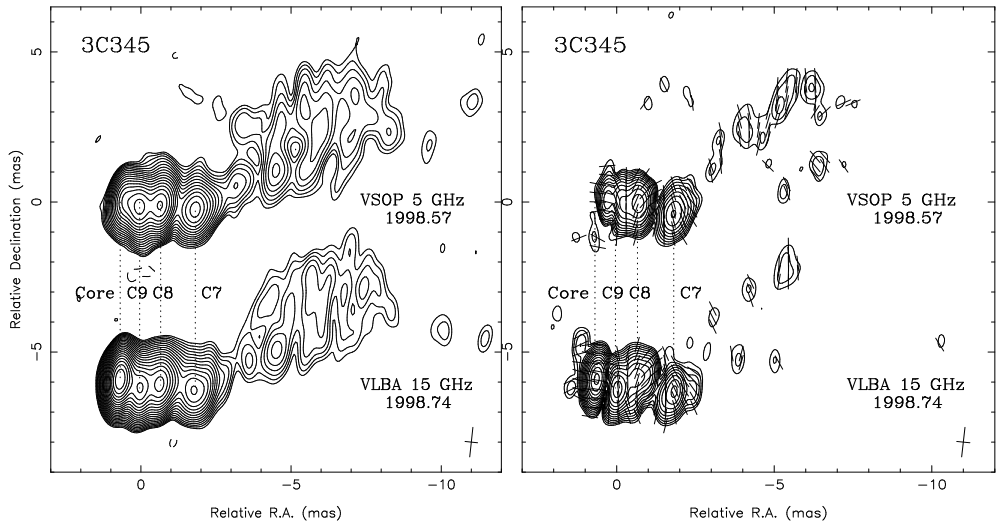}{5.5cm}{0}{63}{63}{-193}{-200}
\vspace{2.75in}
\caption{Matched resolution (as in 3C~279; Fig.~1) total intensity [left 
panel] and polarized intensity with ticks indicating electric field 
direction [right] images of 3C~345. 
\label{fig2}}
\end{figure}

\acknowledgements
The NRAO is operated by Associated Universities, Inc., under cooperative
agreement with the NSF. Space VLBI at NRAO is funded by NASA. The VSOP
project is led by the Japanese ISAS in cooperation with many organizations
and radio telescopes around the world. Radio astronomy at Brandeis is
supported by the NSF.

\end{document}